\documentclass{IEEEtran}
\usepackage{textcomp}

\usepackage{hyperref}
\usepackage{xcolor,soul,framed} 
\colorlet{shadecolor}{yellow}
\newif\ifrevmarks
\revmarksfalse
\definecolor{revA}{RGB}{204,0,0}  
\definecolor{revB}{RGB}{0,0,204}  
\definecolor{revC}{RGB}{0,128,0}  
\definecolor{revD}{RGB}{153,0,153}  
\DeclareRobustCommand{\revA}[1]{{\ifrevmarks\color{revA}\fi#1}}
\DeclareRobustCommand{\revB}[1]{{\ifrevmarks\color{revB}\fi#1}}
\DeclareRobustCommand{\revC}[1]{{\ifrevmarks\color{revC}\fi#1}}
\DeclareRobustCommand{\revD}[1]{{\ifrevmarks\color{revD}\fi#1}}
\usepackage{graphicx}
\graphicspath{{../pdf/}{../jpeg/}}
\DeclareGraphicsExtensions{.pdf,.jpeg,.png}

\usepackage{enumitem}   %
\ifdefined\labelindent\else\newlength\labelindent\fi 
\usepackage{amsmath,amssymb,amsfonts,empheq,amsthm}
\usepackage{mathrsfs}

\usepackage{cite}

\theoremstyle{remark}
\newtheorem{remark}{\indent Remark}

\theoremstyle{definition}
\newtheorem{definition}{Definition}
\newtheorem{lemma}{Lemma}

\newtheorem{theorem}{Theorem}

\newcommand{\mrm}[1] {\mathrm{#1}}
\newcommand{\mbf}[1] {\mathbf{#1}}

\newcommand{\re}[1] {\mathrm{Re}\{#1\}}

\newcommand{\diag}[1] {\mathrm{diag}\{#1\}}

\newcommand{\ml}{\mathrm{l}}

\newcommand{\mP}{\mathrm{P}}
\newcommand{\mI}{\mathrm{I}}
\newcommand{\md}{\mathrm{d}}
\newcommand{\mr}{\mathrm{r}}
\newcommand{\mo}{\mathrm{o}}
\newcommand{\mpv}{\mathrm{pv}}

\usepackage{bm}

\usepackage{array}
\usepackage{mdwmath}
\usepackage{mdwtab}
\usepackage{eqparbox}
\usepackage{url}
\usepackage{threeparttable}
\usepackage{booktabs}
\usepackage{multirow} 
\usepackage{algorithmic}
\usepackage{algorithm}
\usepackage{color}
\usepackage{soul}

\begin{document}
\title{{\fontsize{20pt}{18pt}\selectfont Passivity-Based Decentralized Stability \revA{Criterion} and Active-Damping Design for DC Microgrids}}
\author{Zelin~Sun, 
    Shanshan~Jiang,
    Xiaoyu~Peng,
    Xiuqiang~He,
    Hua~Geng,~\IEEEmembership{Fellow,~IEEE}
  }


\IEEEaftertitletext{\vspace{-3ex}}
\maketitle
\begin{abstract}
\revB{Interactions among converter-interfaced devices can destabilize DC microgrids (DCMGs).
In practice, detailed device models often cannot be disclosed to the system operator for centralized stability analysis, and only some device controllers can be redesigned.
This paper develops an output-feedback passivity (OFP)-based decentralized small-signal stability criterion for DCMGs with both fixed-controller and tunable-controller devices.
By incorporating the certified OFP indices of fixed-controller devices into an equivalent network, the criterion quantifies a locally verifiable OFP requirement for each tunable-controller device.
Full-order analysis further reveals that the finite current-loop bandwidth imposes upper bounds on the OFP indices of droop-controlled converters, which can remain negative for any PI gains despite positive predictions based on reduced-order analysis.
Guided by the criterion and these bounds, an OFP-oriented active damping (OFP-AD) strategy is proposed for these converters. The proposed strategy improves the OFP index toward the required value while preserving the cascaded PI control and droop characteristics, using output-capacitor-current feedback.
Case studies on a 39-node DCMG validate the proposed criterion and OFP-AD.}
\end{abstract}


%
\IEEEpeerreviewmaketitle


%
\vspace{-1\baselineskip}
\section{Introduction}
DC microgrids (DCMGs) interconnect distributed sources
and loads via power electronic converters.
Although individual converters may achieve satisfactory
regulation when considered separately, their electrical
interactions can destabilize the interconnected system
\cite{liu2022complete}.
Conventional eigenvalue- and impedance-based analyses
can assess interconnection stability. However, the resulting stability conditions may couple
the models or parameters of multiple devices.
Device assessment and local controller design instead
call for requirements that can be checked separately for each device and whose joint satisfaction guarantees system small-signal stability.
This device-level verification is particularly
valuable when detailed device models are not
disclosed to the system operator for confidentiality
reasons \cite{haberle2026next,peng2026compositional}.

Existing studies have developed decentralized
stability criteria based on local device input-output properties to obtain such device-level requirements
\cite{haberle2026next,peng2026compositional,wang2017decentralized,chevalier2021network,dey2022passivity,laib2023decentralized,huang2024gain,yang2025compositionalPartI,yang2026compositionalPartII, yang2019distributed, haberle2026decentralized,gorbunov2027dynamic}.
Among these approaches, this paper focuses on
criteria based on port passivity, namely passivity
as viewed from a device's electrical connection
to the network.
Output-feedback passivity (OFP) indices quantify
device passivity excesses and shortages \cite{peng2026compositional,bao2007process,peng2026positive,he2024passivity,sun2025decentralized}. These indices provide a common measure for
comparing the passivity required for system
stability with that provided by a device, and
for evaluating the effects of controller modifications.
For DCMGs, device OFP indices have been coordinated with the network to determine local
requirements for system stability \cite{sun2025decentralized_Conference,sun2025decentralized,sun2026decentralized}. The positive-damping-region method supports local
certification of OFP indices from device frequency
responses \cite{peng2026positive}. 
For DCMGs where only some device controllers
can be redesigned, this paper examines
how to obtain OFP requirements for system
stability, how \revB{the dynamics of conventionally controlled devices limit achievable OFP indices, and how local controller modification can improve these indices}.

Devices whose controllers remain unchanged are
referred to as fixed-controller devices, whereas
those available for redesign are referred to as
tunable-controller devices.
This distinction accounts for practical restrictions
on controller modification
\cite{haberle2021control,haberle2025dynamic}. The certified OFP indices of fixed-controller
devices have to be treated as given in passivity-based decentralized analysis. The challenge is to determine whether
system stability can be certified by imposing
sufficient OFP requirements on the tunable devices
and, if so, to derive these requirements from the
network and the fixed-device indices.

The validity of stability assessment also depends on
the device models used to certify OFP indices. Several studies employ reduced-order device models that neglect fast inner-loop dynamics \cite{jiang2025mode,jiang2025state,sun2025decentralized_Conference,sun2025decentralized,sun2026decentralized,yang2019distributed,he2024passivity,haberle2026decentralized,peng2026compositional}. Singular perturbation theory supports this approximation for slow dynamics under suitable assumptions \cite{khalil2002nonlinear}, but does not guarantee preservation of port passivity. Recent work has also identified limitations of passivity certification for higher-fidelity converter models \cite{peng2026relative}. In particular, finite-bandwidth current-loop dynamics may invalidate OFP indices predicted under ideal current tracking. OFP-based stability certificates obtained using
reduced-order models may not apply to the full-order model of the same microgrid.
The full-order model may reveal instability
even when the reduced- and full-order device-port
responses closely agree at low frequencies. It is thus necessary to \revB{quantify how the controller structure and gains} constrain the
\revD{OFP indices} in full-order models.
Understanding these limitations supports reliable
stability certification and identifies requirements
that cannot be met by gain tuning alone.

When existing controllers cannot provide the required indices, local control modification can improve device passivity and help satisfy
the decentralized stability condition. Passivity-based methods have been developed for converter stabilization \cite{li2025passivity,cucuzzella2019robust}. 
However, passivity established with respect to
a converter control input does not automatically
imply electrical port passivity.
Port passivization has been achieved through voltage PI control augmented with high-pass filtering and output-current feedback \cite{yu2018passivity}, and through state feedback combined with feedforward compensation \cite{kato2021passivity}. For the control-design problem considered here, the objective is not to render a converter passive, but to improve its OFP index toward a quantitative requirement determined by the interconnected system. The design challenge is to achieve this improvement while preserving the cascaded voltage-current PI structure and steady-state characteristics. Such a design can be integrated into existing
converter controllers without requiring a new
regulation architecture.

\revB{To address these issues, this paper first develops a decentralized stability criterion that specifies OFP requirements for tunable-controller devices. It then examines whether conventionally controlled devices can meet these requirements and designs active damping for those that cannot. The main contributions are as follows:
\begin{enumerate}[leftmargin=*]
\item \textit{Decentralized stability criterion with coexisting fixed- and tunable-controller devices:} Certified OFP indices of fixed-controller devices are incorporated into an equivalent network, from which a locally verifiable OFP requirement for tunable-controller devices is quantified. Meeting this requirement, together with a network damping condition, certifies small-signal stability. The requirement can also be specified uniformly over a prescribed operating set.

\item \textit{Finite-bandwidth current-loop limitations on device OFP indices:} Analytical upper bounds on \revD{OFP indices} are derived from full-order models of cascaded PI-controlled converters. The bounds identify cases in which gain tuning alone, including widening the current-loop bandwidth, cannot yield nonnegative indices despite positive reduced-order predictions.

\item \textit{OFP-oriented active damping (OFP-AD) for droop-controlled converters:} When the existing controller cannot meet the requirement in 1), OFP-AD supplements it with output-capacitor-current feedback. It improves the full-order OFP index toward the requirement while preserving the cascaded PI control and droop characteristics. The damping gain is selected using necessary conditions from the analysis in 2), together with implementation constraints.
\end{enumerate}}

The remainder of this paper is organized as follows. Section~\ref{section:modeling} introduces the DCMG model. Section~\ref{section:decentralized_condition} develops the decentralized stability condition and local certification procedure. Section~\ref{section:device_passivity} analyzes the OFP indices of conventionally controlled devices. Section~\ref{section:controller_redesign} \revB{proposes the OFP-AD strategy}, followed by case studies and conclusions.




\section{DC Microgrid Modeling}\label{section:modeling}
Consider a DCMG comprising $N$ nodes interconnected through a resistive network, with node set $\mathscr V=\{1,\dots,N\}$. Each node is interfaced with an energy storage system (ESS), a photovoltaic (PV) unit, or a constant power load (CPL). An ESS may supply or absorb power. For node $k\in\mathscr{V}$, $u_k$ and $i_k$ denote the node voltage and the current injected by the device into the network, respectively. The small-signal device port impedance around a steady-state operating point $(i_k^*,u_k^*)$ is defined in the frequency domain:
\vspace{-0.5\baselineskip}
\begin{equation}\label{eq:source device port impedance}
Z_k(s):=\frac{\Delta u_k(s)}{-\Delta i_k(s)},
\end{equation}
where  $-\Delta i_k$ is the small-signal incremental current entering the device and $\Delta u_k$ is the incremental port voltage.

Source converters are modeled in a general form so that the stability condition in Section~\ref{section:decentralized_condition} is independent of specific controller realization. CPL models are specified directly, since their controllers are usually unavailable for modification in practice and are kept fixed in this paper. Continuous-time averaged converter models are used, and small-signal analysis is conducted around unsaturated equilibria. The node subscript $k$ is omitted in device-level derivations when unambiguous.

\vspace{-0.5\baselineskip}
\subsection{Transmission Network Model}
The transmission network is modeled as a linear resistive grid with the admittance matrix $Y_\mrm{net}$ \cite{liu2022complete}. Its node currents $\bm i=[i_1,\dots,i_N]^\top$ and voltages $\bm u=[u_1,\dots,u_N]^\top$ satisfy:
\vspace{-0.5\baselineskip}
\begin{equation}\label{eq:network Y}
\bm i =Y_\mrm{net} \bm u.
\vspace{-0.5\baselineskip}
\end{equation}

\vspace{-0.5\baselineskip}
\subsection{General Source-Device Model}
\begin{figure}[t]
    \centering
\includegraphics[width=0.95\linewidth]{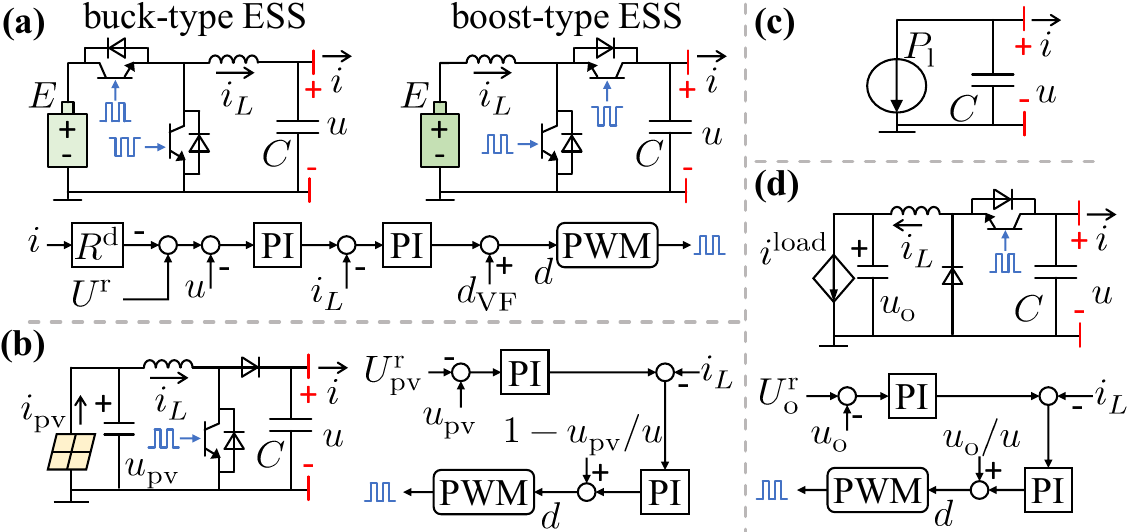}
    \vspace{-1\baselineskip}
    \caption{Port converters and controllers for (a) buck-type or boost-type ESS, (b) PV, (c) quasi-static CPL, and (d) detailed converter-interfaced CPL.}
    \label{fg:device}
    \vspace{-1\baselineskip}
\end{figure}

An ESS under voltage-droop control or a PV unit under maximum power point (MPP) voltage reference, as illustrated in Fig.~\ref{fg:device}(a) and (b), is represented by:
\vspace{-0.5\baselineskip}
\begin{equation}\label{eq:general_source_dynamics}
\begin{aligned}
\dot {\bm x}=\bm f(\bm x, u,i),\quad C\dot u=g(\bm x, u,i) - i
\end{aligned}
\vspace{-0.5\baselineskip}
\end{equation}
where $\bm x$ collects all internal states (e.g., inductor currents, controller-integrator states), and $\bm f(\bm x,u,i)$ describes their dynamics. $C$ and $u$ are the port capacitance and voltage. $g(\bm x, u, i)$ is the converter-side current supplied to the port, e.g., \(g=(1-d)i_L\) for a boost converter with duty ratio $d$ and inductor current $i_L$ (see Fig.~\ref{fg:device}(b)). Linearizing \eqref{eq:general_source_dynamics} around an equilibrium $(u^*,\bm x^*,i^*)$, taking Laplace transforms, and eliminating $\Delta\bm x$ yield the port impedance $Z(s)$ in \eqref{eq:source device port impedance}.

\vspace{-0.5\baselineskip}
\subsection{CPL Model}
In this work, we consider two different CPL representations: a quasi-static CPL and a detailed converter-interfaced CPL, as shown in Fig.~\ref{fg:device}(c) and (d), respectively.

\subsubsection{Quasi-static CPL} 
A constant-power load with port capacitance $C$ satisfies:
\vspace{-0.5\baselineskip}
\begin{equation}\label{eq:static CPL}
    C \dot{u} = -P_{\ml}/u - i,
    \vspace{-0.5\baselineskip}
\end{equation}
where $P_\ml > 0$ is the load power. Neglecting $C$ in the limiting case $C \to 0$ gives the ideal static characteristic $-P_\ml = u i$. \revD{Linearizing \eqref{eq:static CPL} around $(u^\ast,i^*)$ yields the port impedance:}
\vspace{-0.5\baselineskip}
\begin{equation}\label{eq:quasi_static_CPL_impedance}
\revD{Z(s)=\frac{1}{Cs-P_{\ml}/(u^\ast)^2}.}
\end{equation}

\subsubsection{Detailed Converter-interfaced CPL} 
A buck converter supplies a resistive load $R_\ml$ and regulates its voltage $u_\mo$ to $U_\mo^\mr$. At equilibrium, voltage regulation maintains the load power $P_\ml=(U_\mo^\mr)^2/R_\ml$ independently of the port voltage. The detailed model retains the converter and controller dynamics:
\vspace{-0.5\baselineskip}
\begin{equation}
\begin{aligned}
C\dot u&=-d i_L-i,
&L\dot i_L &=d u-u_\mo
\\ \dot\eta&= k_\mI^u(U_\mo^\mr - u_\mo),
& \dot\eta^i&= k_\mI^i(i_L^\mr - i_L)
\\ C_\mo\dot u_\mo&=i_L - u_\mo/R_\ml
\end{aligned}
\end{equation}
Here, $L$ and $C_\mo$ are the filter inductance and load-side capacitance, respectively. The cascaded PI controller generates:
\vspace{-0.5\baselineskip}
\begin{equation}\label{eq:CPL_reference_and_feedforward}
i^\mr_L = k_\mP^u (U_\mo^\mr - u_\mo)+\eta, ~~ d =\underbrace{k_\mP^i(i_L^\mr - i_L) + \eta^i}_{d_\mrm{PI}} + \underbrace{u_\mo/u}_{d_\mrm{VF}}
\vspace{-0.5\baselineskip}
\end{equation}
where $k_\mP^u,k_\mI^u$ and $k_\mP^i,k_\mI^i$ are the outer voltage-loop and inner current-loop gains; $\eta$ and $\eta^i$ are their integral states. The terms $d_\mrm{PI}$ and $d_\mrm{VF}$ denote the inner PI output and voltage feedforward, respectively. \revD{This model also takes the general form \eqref{eq:general_source_dynamics}, so its port impedance follows by linearizing it and eliminating the internal states, as done in Section~\ref{section:device_passivity}.}

\vspace{-0.5\baselineskip}
\section{Passivity-Based Decentralized Stability Condition}\label{section:decentralized_condition} 
The DCMG is represented as the device-network feedback interconnection in Fig.~\ref{fg:loop_transformation}(a). Passivity theory provides sufficient stability conditions for such interconnections \revD{in terms of the passivity of the devices and the network} \cite{brogliato2020dissipative}. \revD{This section quantifies device passivity and derives local OFP requirements for tunable-controller devices through an equivalent loop transformation. We first recall the definition of passivity.} Here, $A\succ0$ ($\succeq0$) means that a Hermitian matrix $A$ is positive (semi-)definite, and $\lambda_{\min}(A)$ is its smallest eigenvalue.

\begin{definition}\label{def: passivity}\cite[Th.~2.48]{brogliato2020dissipative}
A $p\times p$ real-rational transfer function $M(s)$ is passive if and only if:
\begin{enumerate}[label={\alph*)}]
\item All poles of $M(s)$ are in $\re{s}\leq0$.
\item $M(j\omega)+M^\top(-j\omega)\succeq 0$ for all real $\omega$ for which $j\omega$ is not a pole of $M(s)$.
\item Any pure imaginary pole $j\omega_0$ of $M(s)$ is a simple pole, and the residue $\lim_{s\rightarrow j\omega_0}(s-j\omega_0)M(s)$ if $\omega_0<+\infty$, or $\lim_{\omega\rightarrow \infty}\frac{M(j\omega)}{j\omega}$ if $\omega_0=\infty$, is positive semi-definite.
\end{enumerate}
\end{definition}

For a scalar device impedance $Z_k(s)$ \revD{with port admittance $Y_k(s):=Z_k^{-1}(s)$}, the inequality in condition b) of Def.~\ref{def: passivity} reduces to \revD{$\re{Z_k(j\omega)}\ge0$, or equivalently,}
\vspace{-0.5\baselineskip}
\begin{equation}\label{eq:passivity real part condition}
\revD{\re{Y_k(j\omega)}} \ge 0.
\end{equation}
Individual devices need not be passive. \revD{For example, the quasi-static CPL in \eqref{eq:quasi_static_CPL_impedance} has $Y_k(s)=C_ks-P_{\ml,k}/(u_k^\ast)^2$, whose negative incremental conductance violates \eqref{eq:passivity real part condition} and produces an unstable pole. The passivity excess or shortage of a device is quantified by the following index.}

\begin{definition}[\revD{OFP index \cite{bao2007process}}]\label{def:OFP_index}
\revD{The \textit{output-feedback passivity (OFP) index} of a scalar device impedance $Z_k(s)$ is}
\begin{equation}\label{eq:OFP_index}
\revD{\bar y_k:=\inf_{\omega\in\mathbb R}\re{Y_k(j\omega)}.}
\end{equation}
\end{definition}
\revD{$\bar y_k$ measures the margin of \eqref{eq:passivity real part condition} in units of conductance. A positive $\bar y_k$ indicates a passivity excess: \eqref{eq:passivity real part condition} still holds after a shunt conductance of up to $\bar y_k$ is removed from the device. A negative $\bar y_k$ indicates a passivity shortage: a shunt conductance of at least $-\bar y_k$ must be added for \eqref{eq:passivity real part condition} to hold. For the quasi-static CPL, $\bar y_k=-P_{\ml,k}/(u_k^\ast)^2<0$.}

\subsection{\revD{Equivalent Loop Transformation}}
\begin{figure}[t]
    \centering
\includegraphics[width=0.95\linewidth]{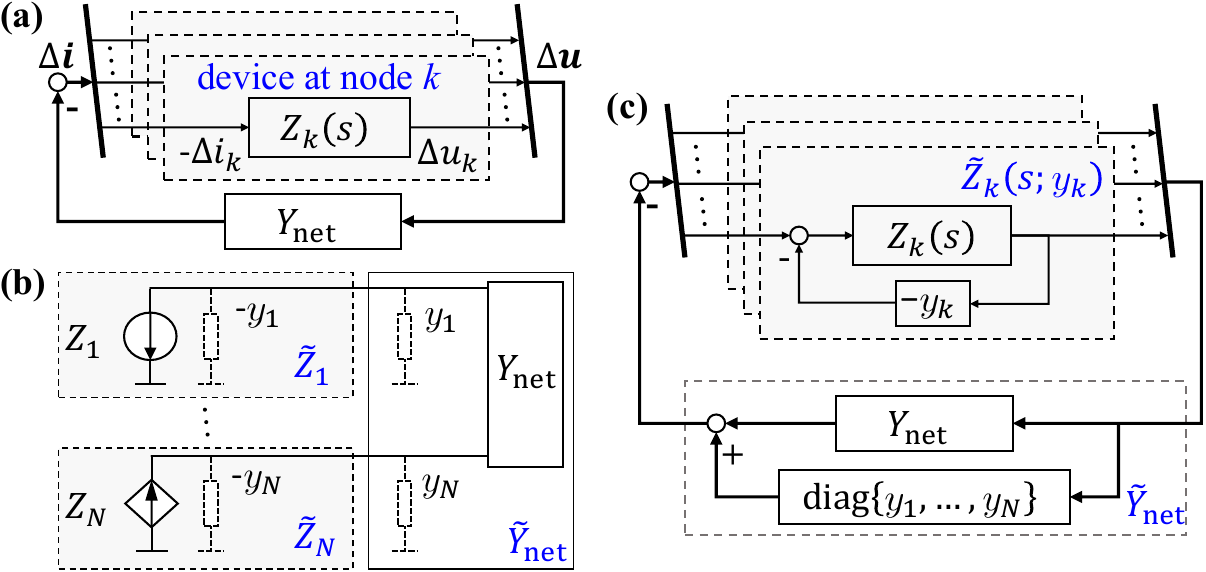}
    \caption{(a) Original device-network interconnection. (b) Virtual shunt-admittance shift. (c) Equivalent transformed interconnection.}
    \label{fg:loop_transformation}
    \vspace{-1\baselineskip}
\end{figure}

\revD{An equivalent loop transformation redistributes passivity excesses and shortages between devices and the network without changing the physical interconnection. On the device side, a constant output feedback with gain $-y_k\in\mathbb R$ yields} the \textit{\textbf{transformed device impedance}}:
\begin{equation}\label{eq:transformed_device_impedance}
\tilde Z_k(s;y_k)=Z_k(s)\left[1-y_kZ_k(s)\right]^{-1}.
\end{equation}
Since $\tilde Z_k^{-1}(s;y_k) = \revD{Y_k(s)}-y_k$, this transformation \textit{virtually} shifts the \revD{shunt conductance of any device} by \(-y_k\), as shown in Fig.~\ref{fg:loop_transformation}(b).

\begin{lemma}\cite{guiver2017transfer} \label{lem:scalar_inverse_passivity}
A nonzero scalar real-rational transfer function \(M(s)\) is passive if and only if \(M^{-1}(s)\) is passive.
\end{lemma}

\revD{By Lem.~\ref{lem:scalar_inverse_passivity}, $\tilde Z_k(s;y_k)$ is passive if and only if $Y_k(s)-y_k$ is passive. By \eqref{eq:OFP_index}, condition b) of Def.~\ref{def: passivity} holds for $Y_k(s)-y_k$ if and only if
 \begin{equation}\label{eq:passivity_index_condition}
     y_k\le\bar y_k\le \re{Y_k(j\omega)},\quad \forall \omega\in\mathbb R.
 \end{equation}
 The OFP index $\bar y_k$ is thus the critical value of the shift, beyond which $\tilde Z_k(s;y_k)$ cannot be passive.}

To preserve closed-loop equivalence, the device-side shift \(-y_k\) is accompanied by the complementary network-side shift \(+y_k\) (see Fig.~\ref{fg:loop_transformation}(c)), yielding the \textit{\textbf{transformed network admittance}}:
\begin{equation}\label{eq:general_transformed_network}
\tilde Y_\mrm{net} = Y_\mrm{net}+\diag{y_1,\dots,y_N}.
\end{equation}
Thus, positive \revD{shifts} add damping to the transformed network, while negative \revD{shifts} reduce it. \revD{In what follows, $y_k\le\bar y_k$ denotes a \textit{certified index}, i.e., a shift for which $\tilde Z_k(s;y_k)$ is passive. A larger certified index increases the corresponding diagonal entry of $\tilde Y_\mrm{net}$ and thus makes the network damping condition less restrictive, regardless of its sign.}

\subsection{Decentralized Stability Condition}
Partition the node set into disjoint sets, $\mathscr V = \mathscr V^{\mrm t} \cup \mathscr V^{\mrm f}$, where: i) \(\mathscr V^{\mrm t}\) contains devices with \textit{tunable controllers}, which may be locally modified to meet an assigned OFP requirement, and ii) \(\mathscr V^{\mrm f}\) contains devices with \textit{fixed controllers}, whose certified OFP indices are supplied as inputs to the system-level assessment.
This partition specifies which device controllers are available for redesign, not dependent on the physical source/load type or the sign of the OFP index. CPL controllers remain fixed in this work. Source controllers may be classified as fixed or tunable according to the coordination scenario.

Ordering the nodes according to \(\mathscr V^{\mrm t}\) and \(\mathscr V^{\mrm f}\) gives: 
\begin{equation}\label{eq:partitioned_modified_network}
\tilde Y_\mrm{net}
=
\begin{bmatrix}
Y^{\mrm{tt}}_\mrm{net}+\mbf y^{\mrm t}
&
Y^{\mrm{tf}}_\mrm{net}
\\
(Y^{\mrm{tf}}_\mrm{net})^\top
&
Y^{\mrm{ff}}_\mrm{net}+\mbf y^{\mrm f}
\end{bmatrix},
\end{equation}
where $\mbf y^{\mrm t} := \diag{y_k^{\mrm t}: k\in\mathscr V^{\mrm t}}, \mbf y^{\mrm f} := \diag{y_k^{\mrm f}: k\in\mathscr V^{\mrm f}}$.
The certified indices \(\mathbf y^{\mathrm f}\) are treated as given, whereas the requirements on \(\mathbf y^{\mathrm t}\) are determined below.

Port impedances describe device input-output behavior and omit internal modes that are not excited by port current perturbations or not visible in the port voltage. Stability of these hidden modes must therefore be checked separately.


\begin{theorem}\label{thm: uniform bound}
Assume that all device modes not represented in the port impedances are stable. The linearized DCMG is stable if:
\begin{enumerate}[label={\roman*)}]
\item\label{con: Device condition} 
\textit{Device passivity:} $\tilde Z_k(s;y_k)$ is passive, $\forall k\in\mathscr V$.
\item\label{con:subnetwork_condition} 
\textit{Subnetwork damping:} The transformed subnetwork associated with the fixed-controller devices retains strictly positive damping: $Y^{\mrm{ff}}_\mrm{net}+\mbf y^{\mrm f}\succ0$. 
\item\label{con:provider_condition}
\textit{Tunable-device OFP requirement:}
The certified indices satisfy: 
\begin{equation}\label{eq:uniform_bound}
y_k^{\mrm t}
\geq
-\lambda_{\min}(\Xi),
\quad
\forall k\in\mathscr V^{\mrm t},
\end{equation}
where $\Xi := Y^{\mrm{tt}}_\mrm{net} - Y^{\mrm{tf}}_\mrm{net} ( Y^{\mrm{ff}}_\mrm{net}+\mbf y^{\mrm f} )^{-1} Y^{\mrm{ft}}_\mrm{net}$. 
\end{enumerate}
\end{theorem}

\textit{Proof:}
Condition~\ref{con: Device condition} guarantees the passivity of all transformed devices. Thus, closed-loop stability can be certified by establishing the passivity of the transformed network \cite{khalil2002nonlinear}. Since \(\tilde Y_\mrm{net}\) is real symmetric, its passivity condition reduces to \(\tilde Y_\mrm{net}\succeq0\). Under condition~\ref{con:subnetwork_condition}, applying the Schur complement to \eqref{eq:partitioned_modified_network} yields: $\tilde Y_\mrm{net}\succeq0 \iff \mbf y^{\mrm t}+\Xi\succeq0$. Condition~\ref{con:provider_condition} implies $\mbf y^{\mrm t} \succeq -\lambda_{\min}(\Xi)I$. Therefore, \(\mbf y^{\mrm t}+\Xi\succeq-\lambda_{\min}(\Xi)I+\Xi\succeq0\). Hence, the transformed network is passive, which establishes the stability of the DCMG.\hfill\(\square\)

\begin{remark}\label{remark:Feasibility of Admittance Shifting}
Condition~\ref{con:subnetwork_condition} is a prerequisite for certification under the selected partition and fixed-device indices: if $Y_{\mathrm{net}}^{\mathrm{ff}}+\mathbf y^{\mathrm f}$ has a negative eigenvalue, no adjustment of $\mathbf y^{\mathrm t}$ can make $\tilde Y_{\mathrm{net}}\succeq0$. 
When all tunable-controller devices are replaced by ideal voltage sources at their equilibrium voltages, Th.~\ref{thm: uniform bound} reduces to Condition~\ref{con:subnetwork_condition}, which thus ensures stability under such ideal voltage support. 
If each fixed device is certified at its OFP index $\bar y_k$ and this index equals $Y_k(0)$, as shown for the PVs and CPLs in Section~\ref{section:device_passivity}, Condition~\ref{con:subnetwork_condition} is also necessary for stability under ideal voltage support. 
In this case, $Y_{\mathrm{net}}^{\mathrm{ff}}+\mathbf y^{\mathrm f}$ is a principal submatrix of the Jacobian $J$ in Appendix~\ref{ap:Equilibrium-Voltage Bounds}, so Condition~\ref{con:subnetwork_condition} holds at any equilibrium with $J\succ0$. Such an equilibrium is the desired operating point of the DCMG and exists if and only if the operating condition lies strictly within the loadability limit. 
Hence, Condition~\ref{con:subnetwork_condition} can fail only at or beyond this limit, where the DCMG loses its desired operating point. Conservative indices $y_k^{\mrm f}<\bar y_k$ make Condition~\ref{con:subnetwork_condition} more restrictive.
\end{remark}

\begin{remark}\label{remark:partition_and_allocation}
The uniform bound \eqref{eq:uniform_bound} is sufficient but potentially conservative. A less restrictive matrix condition $\mathbf y^{\mrm t}+\Xi\succeq0$ allows heterogeneous device requirements. This can accommodate unequal device capabilities, but requires coordinated allocation rather than a single broadcast threshold.
\end{remark}

\subsection{Device OFP Requirements and Local Certification}
\label{subsection:decentralized_certification}
Condition \eqref{eq:uniform_bound} specifies a stability-oriented, model-agnostic minimum OFP requirement for tunable-controller devices. The system operator (SO) determines this requirement from the network and the certified indices of fixed-controller devices, while each tunable-controller device is responsible for achieving and locally verifying compliance.

For a prescribed operating set $\Omega$ of feasible equilibria, the requirements are determined and verified as follows:
\begin{enumerate}[leftmargin=*]
\item The SO specifies the network model, $\Omega$, and device partition. The local operating points required for device certification are provided by the SO or determined locally.
\item Each fixed-controller device reports an index $y_k^\mrm{f}$ for which $\tilde Z_k(s;y_k^{\mrm f})$ is certified passive throughout $\Omega$. \revD{Values close to $\inf_{\Omega}\bar y_k$ are preferred to reduce conservatism.}
\item The SO checks the subnetwork damping condition \ref{con:subnetwork_condition}. If it holds, the SO computes $\Xi$ and issues $-\lambda_{\min}(\Xi)$ as the OFP requirement for tunable-controller devices.
\item Each tunable-controller device certifies an index satisfying \eqref{eq:uniform_bound} throughout $\Omega$, adjusting its controller if necessary.
\item If the network and device conditions hold, Th.~\ref{thm: uniform bound} certifies small-signal stability at every equilibrium in $\Omega$. Otherwise, the operating set, controller designs, or device partition should be reconsidered.
\end{enumerate}

Detailed device models and controller designs remain local; the SO only requires certified fixed-device indices and confirmation of tunable-device compliance.
Device connection or disconnection at existing buses need not change the unreduced network admittance if the buses and lines remain unchanged \cite[Remark~5]{sun2025decentralized}. However, the operating conditions and fixed-device indices may change, so the SO must verify that the conditions of Th.~\ref{thm: uniform bound} remain satisfied. The assessment may be performed before connection and updated when the network, device characteristics, or prescribed operating set changes.

Local OFP certification can be performed analytically or using the positive-damping-region method in \cite{peng2026positive}. Specifically, for an impedance $Z_k(s)$ with no pure imaginary poles, a candidate index \(y_k\) can be certified by the following checks:
\begin{enumerate}[label={\alph*)},leftmargin=*]
\item All poles of $\tilde Z_k(s;y_k)$ lie in the left half-plane. For $y_k\neq0$, this is checked by the Nyquist criterion for $-y_kZ_k(s)$ about $(-1,0)$. For $y_k=0$, \revD{the poles of} $Z_k(s)$ are checked.
\item \revD{The} Nyquist trajectory of $Z_k(j\omega)$ lies in the positive-damping region $\mathcal{P}_{\mathrm{PD}}(y_k)$, which is equivalent to \eqref{eq:passivity_index_condition}:
\begin{equation}
\hspace{-2em}
\mathcal{P}_{\mathrm{PD}}(y_k) :=
\begin{cases}
\{ z\in\mathbb{C} : \left|z-\frac{1}{2y_k}\right|\le\frac{1}{2y_k} \}, & y_k>0, 
\\\{z\in\mathbb C:\re{z}\ge0\},& y_k=0,
\\ \{ z\in\mathbb{C} : \left|z-\frac{1}{2y_k}\right|\ge\frac{1}{-2y_k} \}, & y_k<0.
\end{cases}
\end{equation}
\end{enumerate}
\revD{Examples of $\mathcal{P}_{\mathrm{PD}}(y_k)$ for ESSs are shown in Fig.~\ref{fg:ESS nyquist and PD region}. Candidate indices can be increased toward the critical value $\bar y_k$.}



\section{OFP Analysis of Conventionally Controlled Devices}\label{section:device_passivity}
The stability assessment in Section~\ref{section:decentralized_condition} requires certified device OFP indices. This section analyzes \revD{the OFP indices}
of conventionally controlled devices and derives
analytical upper bounds on these indices.
The bounds reveal how cascaded-PI
controller structure and gains limit \revD{these} indices, and when gain tuning alone cannot meet a target OFP requirement. After specifying the source models, we analyze reduced- and full-order models to assess the effect of current-loop dynamics on device OFP indices.

\subsection{Conventional Cascaded PI-Controlled Source Models}
The general source model in Section~\ref{section:modeling} is instantiated by conventional cascaded PI control. The outer loop generates an inductor-current reference, while the inner loop determines the duty ratio together with voltage feedforward: 
\vspace{-0.5\baselineskip}
\begin{equation}\label{eq:duty ratio}
d=d_0:= d_{\mrm{PI}}+d_{\mrm{VF}},\quad d_{\mrm{PI}}
:= k_\mP^i(i_L^\mr-i_L)+\eta^i,
\end{equation}
where the feedforward term $d_{\mrm{VF}}$ is specified for each converter below. The port capacitance and filter inductance are $C$ and $L$. The voltage-loop gains are $k_\mP^u,k_\mI^u$, and the current-loop gains are $k_\mP^i,k_\mI^i$, with integral states $\eta$ and $\eta^i$, respectively. All capacitances, inductances, and PI gains are positive.

\subsubsection{ESS Model}
An ESS regulates the node voltage via a bidirectional converter under droop control, as shown in Fig.~\ref{fg:device}(a). Battery voltage is represented by a constant voltage source $E>0$ and its dynamics are neglected. For both buck- and boost-type interfaces, the inductor-current reference is
\begin{equation}\label{eq:ESS_current_reference}
i_L^\mr=k_\mP^u(U^\mr-R^\md i-u)+\eta,
\end{equation}
where $R^\md>0$ is the droop coefficient and $U^\mr$ is the voltage reference.

When \(E\) is higher than the network voltage, a buck-type interface is employed. The \textit{buck-type ESS} is linear in its state:
\vspace{-0.5\baselineskip}
\begin{equation}\label{eq:model_buck_ESS}
\begin{aligned}
C\dot u&=i_L-i,
& L\dot i_L&=dE-u,\\
\dot\eta&=k_\mI^u(U^\mr-R^\md i-u),
& \dot\eta^i&=k_\mI^i(i_L^\mr-i_L).
\end{aligned}
\vspace{-0.5\baselineskip}
\end{equation}
The voltage feedforward term is $d_{\mrm{VF}}=u/E$.

When \(E\) is lower than the network-voltage range, a boost-type interface is employed. The full-order nonlinear dynamics of the \textit{boost-type ESS} are:
\vspace{-0.5\baselineskip}
\begin{equation}\label{eq:model_boost_ESS}
\begin{aligned}
C\dot u&=(1-d)i_L-i,
& L\dot i_L&=E-(1-d)u,\\
\dot\eta&=k_\mI^u(U^\mr-R^\md i-u),
& \dot\eta^i&=k_\mI^i(i_L^\mr-i_L).
\end{aligned}
\vspace{-0.5\baselineskip}
\end{equation}
The voltage feedforward term is $d_{\mrm{VF}}=1-E/u$.

\subsubsection{PV Model}
The PV unit regulates the array voltage to an MPP reference by a boost converter, as shown in Fig.~\ref{fg:device}(b), with nonlinear dynamics:
\vspace{-0.5\baselineskip}
\begin{equation}\label{eq:model PV}
\begin{aligned}
C\dot u&=(1-d)i_L-i,
& L\dot i_L&=u_\mpv-(1-d)u,\\
\dot\eta&=k_\mI^u(u_\mpv-U_\mpv^\mr),
& \dot\eta^i&=k_\mI^i(i_L^\mr-i_L),\\
C_\mpv\dot u_\mpv&=i_\mpv-i_L.
\end{aligned}
\vspace{-0.5\baselineskip}
\end{equation}
Here, $i_\mpv$ and $u_\mpv$ are the PV output current and voltage, satisfying the implicit I-V characteristic $f_\mpv^\mrm{IV}(i_\mpv,u_\mpv)=0$. $C_\mpv$ is the capacitor in parallel with the PV array, and $U^\mr_\mpv$ is the MPP reference voltage. The inductor-current reference and voltage feedforward are
\begin{equation}\label{eq:PV_reference_and_feedforward}
i_L^\mr=k_\mP^u(u_\mpv-U_\mpv^\mr)+\eta, \quad d_{\mrm{VF}}=1-u_\mpv/u.
\end{equation}
The MPP reference, irradiance, and temperature are assumed fixed over the time scale of the small-signal analysis. Accordingly, the MPP reference-generation dynamics are excluded.

\subsection{Reduced-Order OFP Indices}
Time-scale-based model reduction is widely used in microgrid stability analysis \cite{jiang2025mode,jiang2025state}. When the fast current-loop dynamics are stable and sufficiently separated from the voltage-loop dynamics, singular perturbation theory supports replacing fast states by quasi-steady relations \cite{khalil2002nonlinear}. The resulting OFP indices provide a comparison for full-order analysis.

For a buck-type ESS, replacing the fast current-loop dynamics of $i_L$ and $\eta^i$ by their quasi-steady relations yields:
\begin{equation}\label{eq:buck_reduced_dynamics}
\begin{aligned}
C\dot u&=i_L-i,
 &0&=dE-u
\\ \dot\eta&=k_\mI^u(U^\mr-R^\md i -u),
& 0&= k_\mI^i(i^\mr_L - i_L)
\end{aligned}
\end{equation}
Linearizing \eqref{eq:buck_reduced_dynamics} around a steady-state point $(i^*,u^*)$ yields the reduced-order impedance:
\begin{equation}
Z^{\mrm{ro}}_{\mrm{buck}}(s)=\frac{(R^\md k_\mP^u + 1)s + R^\md k_\mI^u}{Cs^2+ k_\mP^u s + k_\mI^u}
\end{equation}

Applying the same approximation to the boost-type ESS, PV, and converter-interfaced CPL yields, respectively:
\begin{equation}
Z^{\mrm{ro}}_{\mrm{boost}}(s)=\frac{(E k_\mP^u R^\md + u^*)s+E R^\md k_\mI^u}{C u^* s^2 + (\frac{U^\mr- u^*}{R^\md} +E k_\mP^u)s + E k_\mI^u}
\end{equation}
\begin{equation}\label{eq:reduced order PV model}
Z^{\mrm{ro}}_\mpv = \frac{1}{Cs + \frac{U_\mpv^\mr i_\mpv^*}{(u^*)^2}}    
\end{equation}
\begin{equation}\label{eq:reduced order CPL model}
Z^{\mrm{ro}}_\mrm{cpl} = \frac{1}{Cs  -P_{\ml}/(u^*)^2}  
\end{equation}
The reduced-order PV and converter-interfaced CPL models retain three state variables: $u$, $\eta$, and $u_\mpv$ or $u_\mo$. Their port impedances are nevertheless 1st-order, because \revD{the internal regulation states $\eta$ and $u_\mpv$ or $u_\mo$ are driven by neither $\Delta u$ nor $\Delta i$. These modes are thus not excited by port small-signal perturbations and} do not appear in the port transfer functions. For positive control gains and a nonpositive slope of the PV I-V characteristic, these reduced-order internal regulation modes are Hurwitz and do not introduce hidden instability.

To determine the \revD{OFP indices} of the reduced-order ESSs, consider the general 2nd-order impedance $Z(s)=\frac{a_{1} s+a_{0}}{s^2 + b_{1} s+ b_{0}}$. Its transformed impedance $\tilde{Z}(s;y^\mrm{ro})$ is:
\begin{equation}\label{eq:transformed_second_order_impedance}
\tilde{Z}(s;y^\mrm{ro})= \frac{a_{1} s  + a_{0}}{s^2+(b_{1}-a_{1} y^\mrm{ro})s +b_{0}-a_{0} y^\mrm{ro}}
\end{equation}
Since $a_{1}>0$ and $a_{0}>0$ for the reduced-order ESS models, the \revD{OFP index} is therefore \cite[Exercise~6.7]{khalil2002nonlinear}:
\begin{equation}\label{eq:second_order_OFP_condition}
\bar y^{\mrm{ro}}
=\min\left\{\frac{b_{1}a_{1}-a_{0}}{a_{1}^{2}}, \frac{b_{0}}{a_{0}}\right\}.
\end{equation}
Table~\ref{tab:parameter_mapping} gives the coefficient mapping for different ESS models. Section~\ref{section:case study} will show that positive reduced-order OFP indices are achievable with suitable controller gains.

\begin{table}[t]
\centering
\caption{Mapping of Reduced-Order ESS Parameters to $Z(s)=\frac{a_1 s+a_0}{s^2 + b_1 s+ b_0}$}
\vspace{-1\baselineskip}
\label{tab:parameter_mapping}
\footnotesize
\begin{tabular}{lcccc}
\toprule
ESS & $a_1$ & $a_0$ & $b_1$ & $b_0$ \\
\midrule
Buck-type & $\frac{R^\md k_\mP^u + 1}{C}$ & $\frac{R^\md k_\mI^u}{C}$ & $\frac{k_\mP^u}{C}$ & $\frac{k_\mI^u}{C}$ \\
\addlinespace[4pt]
Boost-type & $\frac{E k_\mP^u R^\md + u^*}{C u^*}$ & $\frac{E R^\md k_\mI^u}{C u^*}$ & $\frac{U^\mr - u^* + E R^\md k_\mP^u}{C u^* R^\md}$ & $\frac{E k_\mI^u}{C u^*}$ \\
\bottomrule
\end{tabular}
\vspace{-1\baselineskip}
\end{table}

The transformed reduced-order PV impedance is:
\begin{equation}
\tilde Z_{\mpv}^{\mrm{ro}}(s;y^\mrm{ro})=\frac{1}{Cs+\frac{U_{\mpv}^{\mr}i_{\mpv}^{\ast}}{(u^\ast)^2}-y^\mrm{ro}}
\end{equation}
The \revD{OFP index} is therefore:
\begin{equation}\label{eq:PV_reduced_OFP}
\bar y^\mrm{ro}=U_{\mpv}^\mr i_{\mpv}^*/{(u^*)^2}\ge0.
\end{equation}
At this \revD{critical} value, $\tilde Z^{\mrm{ro}}_{\mpv}(s;\bar y^\mrm{ro})=\frac{1}{C s}$, which is passive. Since the steady-state power is $P_{\mpv}^*=U_{\mpv}^{\mr}i_{\mpv}^*$, $\bar y^\mrm{ro}$ equals the incremental admittance $\frac{-\partial i}{\partial u}|_{u=u^*}$ of the static constant-power curve $i=\frac{P_{\mpv}^*}{u}$.

The reduced-order converter-interfaced CPL has the same port impedance as the quasi-static CPL \revD{in \eqref{eq:quasi_static_CPL_impedance}}. Its \revD{OFP index therefore equals that of the quasi-static CPL}, i.e.,
\begin{equation}\label{eq:reduced CPL OFP}
\bar y^\mrm{ro}=-P_{\ml}/(u^*)^2\le 0 .
\end{equation}

These models predict positive OFP indices for PV units and suitably controlled ESSs, whereas CPLs exhibit negative indices. Such indices can be used in the decentralized condition in \eqref{eq:uniform_bound} to certify the reduced-order interconnection. 

\subsection{Full-Order OFP Indices and Bounds}\label{subsection:full-order Passivity Degradation}
Relying solely on reduced-order assessment has two limitations:
\begin{enumerate}[label={\roman*)}]
\item The assumption of a ``sufficiently fast'' inner loop does not by itself provide a quantitative, locally verifiable condition on devices under which a reduced-order
stability certificate remains valid for the
full-order interconnection.
\item Neglecting the current-loop dynamics may omit interactions that affect system stability, leading to inaccurate stability predictions.
\end{enumerate}
We therefore characterize the full-order OFP
indices and derive upper bounds that reveal
limitations hidden by the reduced-order approximation.

For both the PV and converter-interfaced CPL, the voltage feedforward gives $L\dot i_L=u d_{\mrm{PI}}$, which linearizes to $L\Delta\dot i_L=u^\ast\Delta d_{\mrm{PI}}$ since $d_{\mrm{PI}}^\ast=0$. Neither this equation nor the remaining internal dynamics depend on $\Delta u$ or $\Delta i$, so port perturbations do not excite the internal states. The port impedances thus follow from the linearized port-capacitor equations:
\begin{equation}
Z^{\mrm{fo}}_\mpv = \frac{1}{Cs + \frac{P^*_\mpv}{(u^*)^2}}, \quad Z^{\mrm{fo}}_\mrm{cpl} = \frac{1}{Cs  -\frac{P_\ml^*}{(u^*)^2}},  
\end{equation}
where $P_\mpv^\ast=U_\mpv^\mr i_\mpv^\ast$ and $P_\ml^*=(U_\mo^\mr)^2/R_\ml$. These impedances coincide with their reduced-order expressions in \eqref{eq:reduced order PV model} and \eqref{eq:reduced order CPL model}, so their \revD{OFP indices} remain those in \eqref{eq:PV_reduced_OFP} and \eqref{eq:reduced CPL OFP}. At a given equilibrium, these indices depend on the operating power and port voltage, rather
than control gains. The gains must nevertheless make the internal regulation matrices Hurwitz to exclude instability hidden from the port transfer functions.

For the ESS models, however, the retained current-loop dynamics can degrade the OFP index. Their OFP limitations are examined using the port admittance $Y^\mrm{fo}(s):= [Z^\mrm{fo}(s)]^{-1}$. \revD{Since \eqref{eq:OFP_index} takes the infimum over all frequencies, the OFP index is bounded above by the high-frequency conductance:}
\begin{equation}\label{eq:OFP_high_frequency_upper_bound}
\revD{\bar y \leq \liminf_{\omega\rightarrow\infty} \re{Y^\mrm{fo}(j\omega)}.}
\end{equation}

The high-frequency bound can rule out a target OFP index, but satisfying it alone does not certify that index. The remaining frequencies and the pole and residue conditions must also be checked.

\subsubsection{Buck-type ESS} 
The full-order admittance satisfies:
\begin{equation}\label{eq:buck_high_frequency_real_part}
\re{Y^\mrm{fo}_{\mrm{buck}}(j\omega)}=-\frac{ECR^{\md}k_{\mP}^{u}k_{\mP}^{i}}{L}+O(\omega^{-2}), ~ \omega\rightarrow\infty.
\end{equation}
For positive controller gains, the high-frequency real part of $Y^\mrm{fo}_{\mrm{buck}}(j\omega)$ is negative. \revD{By \eqref{eq:OFP_high_frequency_upper_bound},}
\begin{equation}
\bar y \le \lim_{\omega\rightarrow\infty}\re{Y^\mrm{fo}_\mrm{buck}(j\omega)}=\frac{-ECR^\md k_\mP^u k_\mP^i}{L}<0
\end{equation}
This upper bound is negative for every choice of positive control gains. Thus, gain tuning alone cannot yield a nonnegative OFP index, even when the reduced-order model predicts a positive index. \revB{Since $L\dot i_L=Ed_\mrm{PI}$, the proportional gain alone yields a first-order current loop with bandwidth $\omega_\mrm{c}^i:=Ek_\mP^i/L$, and the bound above equals $-R^\md k_\mP^u C\omega_\mrm{c}^i$. Widening the current-loop bandwidth through $k_\mP^i$ thus} makes the upper bound more negative rather than removing this limitation. \revB{Neglecting the current-loop dynamics, as in the reduced-order model, corresponds to $\omega_\mrm{c}^i\to\infty$. A larger $\omega_\mrm{c}^i$ indeed makes the full-order admittance agree with the reduced-order one over a wider frequency range, but above this range its real part still turns negative and approaches $-R^\md k_\mP^u C\omega_\mrm{c}^i$. \revD{Since $\bar y\le\re{Y^\mrm{fo}(j\omega)}$ over all frequencies,} the reduced-order index is not recovered even as $\omega_\mrm{c}^i\to\infty$.} Fig.~\ref{fg:ESS nyquist and PD region}(a) shows the high-frequency negative real part of $Y^\mrm{fo}_{\mrm{buck}}(j\omega)$, where the parameters and equilibrium are from the Buck-type ESS at node~1 in Case~1 (Subsection~\ref{subsection:case1}). The local procedure in Subsection~\ref{subsection:decentralized_certification} certifies \revD{the index} $\bar y=-0.05712~\Omega^{-1}$, with the corresponding impedance trajectory and positive-damping region shown in Fig.~\ref{fg:ESS nyquist and PD region}(c).

\subsubsection{Boost-type ESS} 
\begin{figure}[t]
    \centering
\includegraphics[width=0.9\linewidth]{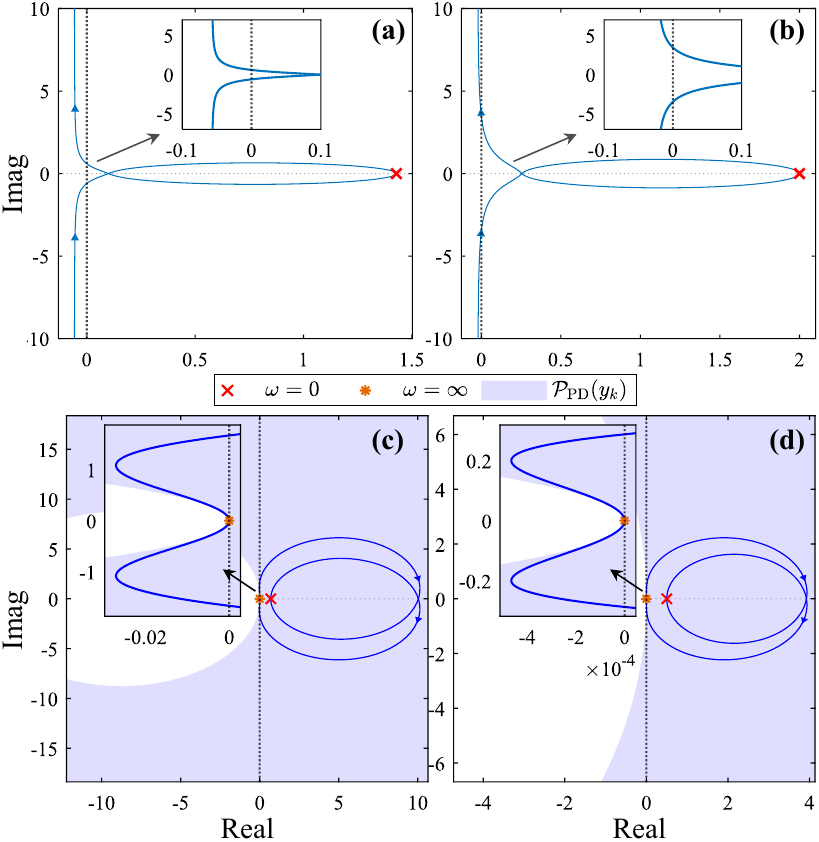}
\vspace{-0.5\baselineskip}
    \caption{Full-order ESS port responses under cascaded PI control. Admittance Nyquist plots of the (a) buck-type ESS and (b) boost-type ESS. Impedance Nyquist plots and positive-damping regions of the (c) buck-type ESS ($y=\bar y=-0.05712~\Omega^{-1}$) and (d) boost-type ESS ($y=\bar y=-0.0242~\Omega^{-1}$).}
    \label{fg:ESS nyquist and PD region}
    \vspace{-1\baselineskip}
\end{figure}

The real part of its admittance satisfies:
\begin{equation}\label{eq:boost full order admittance}
\re{Y^\mrm{fo}_\mrm{boost}(j\omega)}=\frac{a (i_L^*)^2 + b i_L^* + c}{L(1-R^\md k_\mP^i k_\mP^u i_L^*)^2} + O(\omega^{-2}),~\omega\rightarrow\infty
\end{equation}
where
\begin{equation}
\!\!\!\!\!
\begin{aligned}
a ={}& L R^\md k_\mP^u k_\mP^i[k_\mP^u k_\mP^i -E/(u^*)^2], \quad c = -C ER^\md k_\mP^i k_\mP^u  
\\ b={}&LE/(u^*)^2+(CL R^\md k_\mP^u k_\mI^i  + CL R^\md k_\mP^i k_\mI^u- L k_\mP^u k_\mP^i)
\\  &- C R^\md k_\mP^u (k_\mP^i)^2 u^* 
\end{aligned}
\end{equation}
The boost-type ESS has an operating-point-dependent OFP upper bound. Since $c<0$, the high-frequency real part of $Y^\mrm{fo}_\mrm{boost}(j\omega)$ is negative in a neighborhood of \(i_L^\ast=0\), yielding:
\begin{equation}
\bar y\le\lim_{\omega\rightarrow\infty}\re{Y^\mrm{fo}_\mrm{boost}(j\omega)}<0.
\end{equation}
Gain tuning alone thus cannot eliminate a negative OFP index at an operating point near \(i_L^\ast=0\). \revB{Here, $L\dot i_L=ud_\mrm{PI}$, so the proportional gain alone gives the current-loop bandwidth $\omega_\mrm{c}^i:=u^\ast k_\mP^i/L$ around $u^\ast$. The limit at $i_L^\ast=0$ then equals $c/L=-(E/u^\ast)R^\md k_\mP^u C\omega_\mrm{c}^i$, exhibiting the same bandwidth dependence as the buck-type ESS.}
Fig.~\ref{fg:ESS nyquist and PD region}(b) shows a negative high-frequency real part, where the parameters and equilibrium are from the Boost-type ESS at node~13 in Subsection~\ref{subsection:case1} Case~1. The local procedure in Subsection~\ref{subsection:decentralized_certification} certifies \revD{the index} $\bar y=-0.0242~\Omega^{-1}$. The impedance trajectory and $\mathcal{P}_{\mrm{PD}}(-0.0242)$ are shown in Fig.~\ref{fg:ESS nyquist and PD region}(d).

\begin{remark}
The equality between the reduced- and full-order PV and CPL impedances relies on the specified voltage feedforward in \eqref{eq:PV_reference_and_feedforward} and \eqref{eq:CPL_reference_and_feedforward}. Without this term (i.e., $d=d_\mrm{PI}$), their full-order port impedances differ from the reduced-order expressions. For the PV, $\lim_{\omega\to\infty}\re{Y^\mrm{fo}_\mpv(j\omega)}=0$, which rules out a positive OFP index by \revD{\eqref{eq:OFP_high_frequency_upper_bound}}. For the converter-interfaced CPL, the low-frequency expansion as $\omega\to0$ gives $\re{Y^\mrm{fo}_\mrm{cpl}(j\omega)}= -\frac{P_\ml}{(u^\ast)^2} -\frac{2P_\ml} {(u^\ast)^3k_\mI^i k_\mI^u}\omega^2 +O(\omega^4)$, the negative quadratic coefficient making its OFP index below $-P_{\ml}/(u^*)^2$.
\end{remark}

The analysis provides OFP indices and bounds for assessing conventionally controlled devices. Reduced-order indices that \revD{exceed the full-order OFP indices} cannot be used either as fixed-device certificates or to demonstrate tunable-device compliance. For an ESS assigned to the tunable-controller set, an upper bound below its required index rules out compliance with the existing controller. The structure-dependent limitations identified show when gain tuning alone is insufficient. Section~\ref{section:controller_redesign} \revB{proposes the OFP-AD strategy} to improve the full-order OFP indices of tunable ESSs.

\section{\revB{OFP-Oriented Active-Damping Control}}
\label{section:controller_redesign}
For an ESS $k\in\mathscr V^{\mrm t}$, controller modification is needed when its existing controller cannot provide the OFP index required by \eqref{eq:uniform_bound}. This section \revB{proposes OFP-oriented active damping (OFP-AD), which uses output-capacitor-current feedback} to increase the achievable OFP indices while retaining cascaded PI control. To address the high-frequency limitations identified in Section~\ref{section:device_passivity}, the original duty-ratio command $d_0$ in \eqref{eq:duty ratio} is augmented with feedback of $i_C=C\dot u$. This realizes voltage-derivative feedback using measured inductor and port currents, without directly measuring $i_C$ or explicit voltage differentiation. \revC{Alternatively, $i_C$ can be estimated from the DC voltage, the modulation signal, and the load current \cite{geng2011active}.} Since $i_C^\ast=0$, the original droop equilibrium is preserved.

The following analysis derives implementation constraints and necessary high-frequency conditions for the target index $y$, with $Y^{\mathrm{ad}}(s)$ denoting the admittance under the augmented controller. These conditions guide parameter selection but do not guarantee compliance, which must be verified using the OFP-certification procedure in Subsection~\ref{subsection:decentralized_certification}.

\vspace{-0.5\baselineskip}
\subsection{Active-Damping Control for Buck-Type ESS}
\label{subsection:buck_active_damping}
For the buck-type ESS, the inductor equation $L\dot i_L=dE-u$ gives a constant duty-ratio-to-inductor-voltage gain \(E>0\). The duty-ratio command is augmented as:
\begin{equation}\label{eq:buck_active_damping_control}
d=d_0-k_{\mathrm{ad}}(i_L-i), \quad k_{\mathrm{ad}}>0,
\vspace{-0.5\baselineskip}
\end{equation}
where \(i_L-i=i_C=C\dot u\).

When the port voltage rises, the added damping term reduces the duty ratio. Its contribution lowers the inductor voltage and hence reduces \(\dot i_L\), opposing the capacitor-charging transient. The resulting high-frequency conductance is:
\begin{equation}\label{eq:buck_active_damping_high_frequency}
\lim_{\omega\rightarrow\infty}
\re{Y^{\mathrm{ad}}_{\mathrm{buck}}(j\omega)}
=\frac{CE}{L}\left(k_{\mathrm{ad}}-R^{\mathrm d}k_{\mathrm P}^{u}k_{\mathrm P}^{i}\right).
\end{equation}
Compared with \eqref{eq:buck_high_frequency_real_part}, OFP-AD adds \(CEk_{\mathrm{ad}}/L>0\) to the high-frequency conductance. \revD{By \eqref{eq:OFP_high_frequency_upper_bound}, achieving $\bar y\ge y$ for a target index $y$ requires}:
\vspace{-0.5\baselineskip}
\begin{equation}\label{eq:buck_active_damping_provider_condition}
k_{\mathrm{ad}} \geq R^{\mathrm d}k_{\mathrm P}^{u}k_{\mathrm P}^{i} +\frac{L}{CE}y.
\end{equation}

\subsection{Active-Damping Control for Boost-Type ESS}
\label{subsection:boost_active_damping}
For the boost-type ESS, the capacitor current is
\vspace{-0.5\baselineskip}
\begin{equation}\label{eq:boost_output_capacitor_current}
i_C=(1-d)i_L-i=C\dot u.
\vspace{-0.5\baselineskip}
\end{equation}
Unlike the buck converter, the duty ratio directly affects $i_C$, with an incremental gain $-i_L^\ast$ that changes sign between charging and discharging. The effect of capacitor-current feedback must thus be evaluated for both power-flow directions.

Consider the capacitor-current feedback:
\vspace{-0.5\baselineskip}
\begin{equation}\label{eq:boost_actual_current_damping}
d=d_0-k_{\mathrm{ad}}i_C, \qquad k_{\mathrm{ad}}>0.
\end{equation}
Since \(i_C\) depends on the final duty ratio, this relation is implicit. Substituting \eqref{eq:boost_output_capacitor_current} gives the explicit implementation:
\begin{equation}\label{eq:boost_explicit_actual_current_damping}
d=d_0-\frac{k_{\mathrm{ad}}}{1-k_{\mathrm{ad}}i_L}[(1-d_0)i_L-i]
\end{equation}
To avoid a singular denominator, impose
\begin{equation}\label{eq:boost_damping_well_posedness}
1-k_{\mathrm{ad}}i_L\geq\varepsilon_d,
\qquad 0<\varepsilon_d<1,
\end{equation}
over the prescribed current envelope. For \(i_L\leq i_{L,\max}\), with \(i_{L,\max}>0\), a sufficient constraint is $k_{\mathrm{ad}} \leq (1-\varepsilon_d)/i_{L,\max}$.

The high-frequency expansion of the resulting admittance begins with
\vspace{-0.5\baselineskip}
\begin{equation}\label{eq:boost_admittance_leading_term}
Y^{\mathrm{ad}}_{\mathrm{boost}}(s)=\frac{C(1-k_{\mathrm{ad}}i_L^\ast)}{1-R^{\mathrm d}k_{\mathrm P}^{i}k_{\mathrm P}^{u}i_L^\ast}s+O(1).
\end{equation}
By Lem.~\ref{lem:scalar_inverse_passivity}, \revD{achieving $\bar y\ge y$} requires \(Y^{\mathrm{ad}}_{\mathrm{boost}}(s)-y\) to be passive. Its coefficient of $s$ must therefore be nonnegative, as required by the residue condition at infinity in Def.~\ref{def: passivity}. Together with \eqref{eq:boost_damping_well_posedness}, this requires:
\begin{equation}\label{eq:boost_PI_leading_coefficient_condition}
1-R^{\mathrm d}k_{\mathrm P}^{i}k_{\mathrm P}^{u}i_L^\ast>0.
\end{equation}

Using the coefficients \(a,b,c\) from \eqref{eq:boost full order admittance}, the high-frequency conductance is
\begin{equation}\label{eq:boost_actual_damping_high_frequency}
\begin{aligned}
\lim_{\omega\rightarrow\infty}
\re{Y^{\mathrm{ad}}_{\mathrm{boost}}(j\omega)}=\frac{a(i_L^\ast)^2+b i_L^\ast+c+Ck_{\mathrm{ad}}\chi(u^\ast,i_L^\ast)}{L\left(1-R^{\mathrm d}k_{\mathrm P}^{i}k_{\mathrm P}^{u}i_L^\ast\right)^2},
\end{aligned}
\end{equation}
where $\chi(u^\ast,i_L^\ast)=E+u^\ast k_{\mathrm P}^{i}i_L^\ast-LR^{\mathrm d}\left(k_{\mathrm P}^{i}k_{\mathrm I}^{u}+k_{\mathrm I}^{i}k_{\mathrm P}^{u}\right)(i_L^\ast)^2$.
The term $C k_{\mathrm{ad}}\chi$ represents the added high-frequency contribution. To make this contribution positive throughout the prescribed charging and discharging range, we impose:
\begin{equation}\label{eq:boost_positive_damping_condition}
\chi(u^\ast,i_L^\ast)>0.
\end{equation}
Under this design condition \eqref{eq:boost_positive_damping_condition}, \revD{\eqref{eq:OFP_high_frequency_upper_bound}} gives the necessary gain condition for the target index \(y\):
\begin{equation}\label{eq:boost_damping_gain_lower_bound}
k_{\mathrm{ad}}
\geq
\frac{
y L\left(
1-R^{\mathrm d}k_{\mathrm P}^{i}k_{\mathrm P}^{u}i_L^\ast
\right)^2
-a(i_L^\ast)^2-b i_L^\ast-c}
{C\chi(u^\ast,i_L^\ast)}.
\end{equation}

At the charging-discharging transition, \(i_L^\ast=0\) and $\chi(u^\ast,0)=E>0$, giving
\begin{equation}\label{eq:boost_zero_current_damping_condition}
\lim_{\omega\rightarrow\infty}
\operatorname{Re}\{Y^{\mathrm{ad}}_{\mathrm{boost}}(j\omega)\}
=
\frac{CE}{L}
\left(
k_{\mathrm{ad}}
-R^{\mathrm d}k_{\mathrm P}^{u}k_{\mathrm P}^{i}
\right).
\end{equation}
Thus, the damping contribution remains $CEk_{\mathrm{ad}}/L>0$ at zero current, and the necessary gain bound reduces to \eqref{eq:buck_active_damping_provider_condition}. 

For prescribed PI gains, first check \eqref{eq:boost_PI_leading_coefficient_condition} and \eqref{eq:boost_positive_damping_condition} throughout the operating set. The gain $k_{\mrm{ad}}$ is then selected to satisfy \eqref{eq:boost_damping_gain_lower_bound} at every equilibrium and \eqref{eq:boost_damping_well_posedness} over the current envelope. If these constraints are incompatible, the PI gains or operating set must be reconsidered for this design. 

The proposed \revB{OFP-AD} is one way to improve device OFP indices toward the requirements. Alternative control designs and their comparison are left for future work.

\section{Case Study}\label{section:case study}
The proposed method is validated on a DCMG using the IEEE 39-node network topology \cite{canizares2015benchmark}. Each line resistance is $0.1~\Omega$. Buck-type ESSs are connected at nodes 1, 2, 5, 6, 9, 10, 11; Boost-type ESSs at nodes 13, 14, 16, 17, 19, 22, 28; PVs at nodes 30--39; and CPLs at the remaining nodes. Each PV unit comprises 7 parallel strings with one QJM200-72 module (provided in MATLAB/Simulink) per string. At $25^\circ\mrm{C}$ and $1000~\mrm{W/m^2}$, the MPP power is $200.1048~\mathrm{W}$ per module, giving $P_{\mpv,0}=1400.7336~\mathrm{W}$ per PV unit. Table~\ref{tab:system_parameters} lists the parameters for both cases.

\begin{table}[t]
\centering
\caption{Physical and Control Parameters of the DCMG}
\vspace{-1\baselineskip}
\label{tab:system_parameters}
\footnotesize
\setlength{\tabcolsep}{3pt}
\begin{tabular}{@{}lcccc@{}}
\toprule
Parameter & Buck-type ESS & Boost-type ESS & PV & CPL \\
\midrule
$L$ (mH) & 1 & 2 & 0.5 & 0.5 \\
$C$ (mF) & 2 & 2 & 0.15 & 0.15 \\
$E$ (V) & 200 & 50 & -- & -- \\
Voltage reference (V) & 105 & 105 & 36.12 & 50 \\
$R^\md$ ($\Omega$) & 0.7 & 0.5 & -- & -- \\
$k_\mP^i$ ($\mathrm{A}^{-1}$)
& 0.0012 & 0.0025 & 0.02 & 0.02 \\
$k_\mI^i$ ($\mathrm{A}^{-1}\mathrm{s}^{-1}$)
& 0.0015 & 0.003 & 20 & 20 \\
$k_\mP^u$ ($\mathrm{A}/\mathrm{V}$)
& 0.17 & 0.16 / 0.4 & 1 & 1 \\
$k_\mI^u$ ($\mathrm{A}/(\mathrm{V\,s})$)
& 12 & 25 & 50 & 50 \\
$k_\mrm{ad}$ ($\mathrm{A}^{-1}$)
& 0.0025 & 0.014 & -- & -- \\
\bottomrule
\end{tabular}

\smallskip
\begin{minipage}{\columnwidth}
\footnotesize
Slash-separated values correspond to Cases 1 and 2, respectively; otherwise, the same value applies to both cases. The voltage references are $U^\mr$ for ESSs, $U^\mr_\mpv$ for PVs, and $U^\mr_\mrm{o}$ for CPLs. For the boost ESS at node 13, both cases use $L=0.5~\mathrm{mH}$, $k_\mP^u=0.4~\mathrm{A/V}$, and $k_\mrm{ad}=0.005~\mathrm{A}^{-1}$. The listed damping gains apply when \revB{OFP-AD} is enabled; otherwise, $k_\mrm{ad}=0$. The listed MPP voltage corresponds to $25^\circ\mathrm{C}$ and $1000~\mathrm{W/m^2}$. The CPL resistance $R_\ml=(U^\mr_\mrm{o})^2/P_\ml$. $C_\mpv=C_\mrm{o}=4~\mathrm{mF}$.
\end{minipage}
\vspace{-0.5\baselineskip}
\end{table}

\subsection{Case 1: OFP-Based Stability Assessment and Stabilization}\label{subsection:case1}
This case compares reduced- and full-order OFP-based stability assessments and evaluates stabilization through \revB{OFP-AD}. ESSs and PVs are assigned to the tunable-controller set $\mathscr V^{\mrm t}$, while CPLs constitute the fixed-controller set $\mathscr V^{\mrm f}$. Each PV supplies $P_{\mpv,0}$, and each CPL has a baseline demand of $1900~\mrm{W}$. The certified CPL indices are incorporated into $\Xi$ to determine the OFP requirement $-\lambda_{\min}(\Xi)$. The PV controllers already meet this requirement and remain unchanged. \revB{OFP-AD} is added only to the ESSs.

\begin{figure}[t]
    \centering
\includegraphics[width=0.95\linewidth]{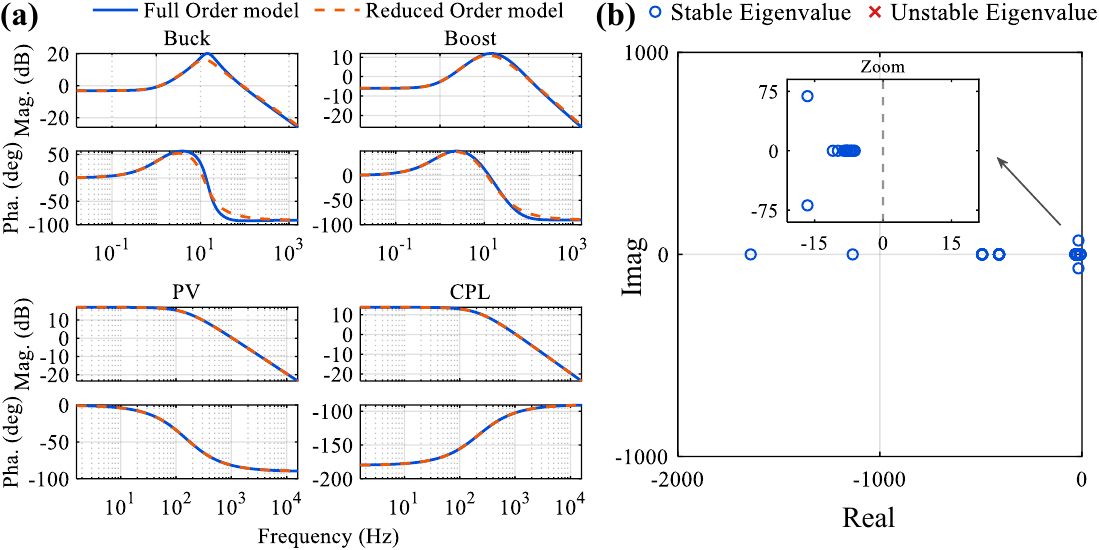}
\vspace{-1\baselineskip}
    \caption{Reduced-order assessment: (a) Bode plots of the reduced-order and full-order device impedances; (b) eigenvalues of the reduced-order DCMG.}
    \label{fg:reduced bode and eigenvalue}
\end{figure}

\begin{figure}[t]
    \centering
\includegraphics[width=0.95\linewidth]{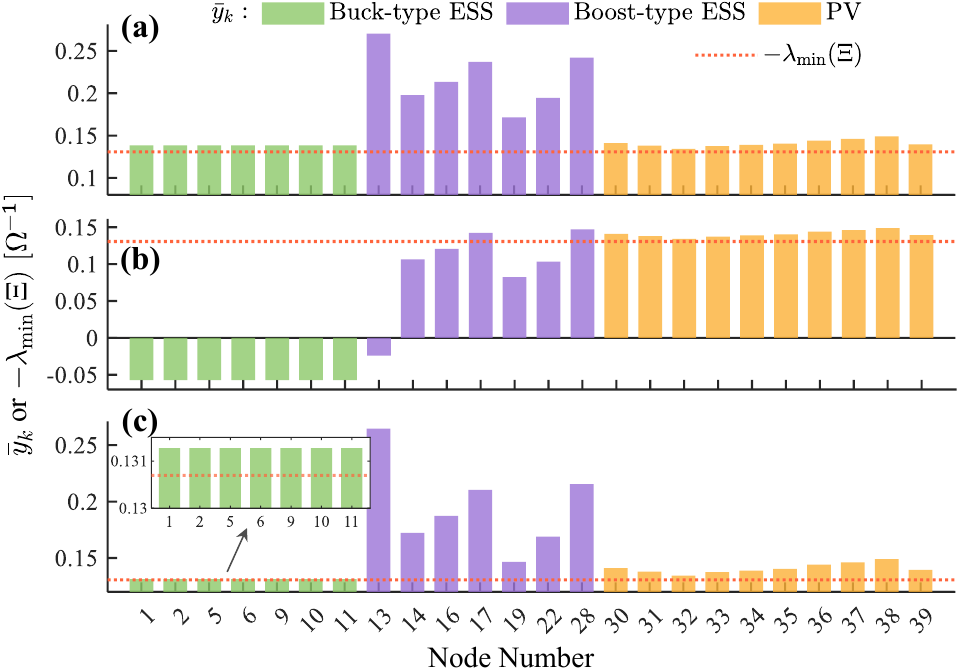}
\vspace{-0.5\baselineskip}
    \caption{Source OFP indices and the OFP requirement $-\lambda_{\min}(\Xi)$: (a) reduced-order models; (b) full-order models without \revB{OFP-AD}; (c) full-order models with \revB{OFP-AD}.}
    \label{fg:case study OFP index}
    \vspace{-0.5\baselineskip}
\end{figure}

Fig.~\ref{fg:reduced bode and eigenvalue}(a) compares the reduced-order and full-order device impedances at nodes 1, 13, 30, and 3. Their similar Bode plots do not imply equivalent passivity properties. The reduced-order source indices satisfy \eqref{eq:uniform_bound} (Fig.~\ref{fg:case study OFP index}(a)), certifying stability of the reduced-order interconnection by Th.~\ref{thm: uniform bound}. Consistently, all reduced-order system eigenvalues lie in the left half-plane (Fig.~\ref{fg:reduced bode and eigenvalue}(b)).

\begin{figure}[t]
    \centering
\includegraphics[width=0.95\linewidth]{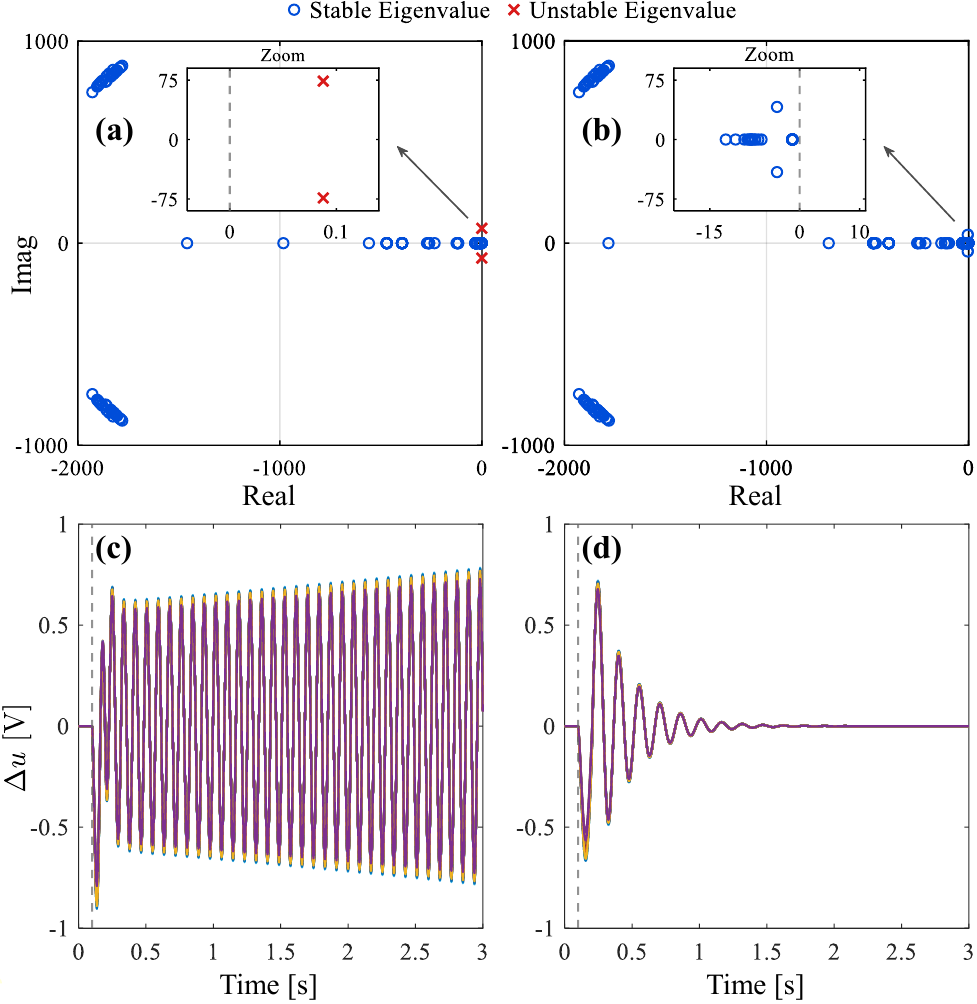}
\vspace{-0.5\baselineskip}
    \caption{Full-order DCMG: eigenvalues (a) without and (b) with \revB{OFP-AD}; voltage responses (c) without and (d) with \revB{OFP-AD}.}
    \label{fg:full order and active damping}
    \vspace{-1\baselineskip}
\end{figure}

Without \revB{OFP-AD}, the full-order indices fail to satisfy the same requirement (Fig.~\ref{fg:case study OFP index}(b)). Failure of this sufficient condition does not assert instability. Here, however, the eigenvalues $0.0878\pm j73.6599$ in Fig.~\ref{fg:full order and active damping}(a) confirm that the full-order system is unstable. A $1\%$ load pulse is applied simultaneously to all CPLs from $t=0.1$ to $0.2~\mathrm{s}$, producing growing voltage oscillations (Fig.~\ref{fg:full order and active damping}(c)).

Adding \revB{OFP-AD to the ESSs} with unchanged PI gains restores compliance with the OFP requirement (Fig.~\ref{fg:case study OFP index}(c)). All full-order eigenvalues in Fig.~\ref{fg:full order and active damping}(b) then lie in the left half-plane, and the voltage responses return to equilibrium following the same pulse (Fig.~\ref{fg:full order and active damping}(d)). This case illustrates the need for full-order compliance assessment and demonstrates stabilization through an additional damping channel.

\subsection{Case 2: Stability Guarantee over an Operating Set}
This case evaluates a uniform OFP requirement over an operating set with varying PV generation and load demand. ESSs are assigned to the tunable-controller set $\mathscr V^{\mrm t}$, while PVs and CPLs constitute the fixed-controller set $\mathscr V^{\mrm f}$. The baseline CPL demand is $P_{\ml,0}=1200~\mathrm{W}$, allowing lighter-load conditions that include ESS charging. The prescribed steady-state powers are:
\begin{equation}
\!\!\!\!P_{\mpv}=\alpha P_{\mpv,0},~ \alpha\in[0.8,1],
\quad
P_\ml=\beta P_{\ml,0},~ \beta\in[0.7,1.3].\!
\end{equation}
The factors $\alpha$ and $\beta$ apply to all PVs and all CPLs, respectively. PV generation is varied via irradiance changes at $25^\circ\mathrm{C}$, with the voltage reference updated from the MPP characteristic. CPL demand is varied via the load resistance.

The min-PV/max-load and max-PV/min-load operating conditions are $(\alpha,\beta)=(0.8,1.3)$ and $(1,0.7)$, respectively. Their high-voltage power-flow solutions, $\bm u^-$ and $\bm u^+$, are shown in Fig.~\ref{fg:case2}(a). The existence and monotonicity results in Appendix~\ref{ap:Equilibrium-Voltage Bounds} establish the \textit{componentwise} bounds
\begin{equation}\label{eq:case2_voltage_bounds}
\bm u^-\leq\bm u^\ast(\alpha,\beta)\leq\bm u^+, \quad \forall\alpha\in[0.8,1],\beta\in[0.7,1.3]
\end{equation}
where $\bm u^\ast(\alpha,\beta)$ denotes the high-voltage power-flow solution. Using these voltage bounds and the full-order OFP expressions in Section~\ref{section:device_passivity}, uniform but conservative fixed-device indices valid throughout this range are selected as:
\begin{equation}\label{eq:case2_fixed_indices}
y_k^{\mrm f}=
\begin{cases}
0.8P_{\mpv,0}/(u_k^+)^2, & \text{PV node},
\\[5pt] -1.3P_{\ml,0}/(u_k^-)^2, & \text{CPL node}.
\end{cases}
\end{equation}
After checking the subnetwork damping condition, the SO incorporates these indices into $\Xi$ and obtains the uniform requirement $-\lambda_{\min}(\Xi)$.

Each ESS is assessed \textit{locally} using its full-order impedances at 1001 equally spaced equilibrium voltages in each interval $[u_k^-,u_k^+]$. The minimum certified OFP index of every ESS exceeds the uniform requirement (Fig.~\ref{fg:case2}(b)), providing numerical support for the stability guarantee. 
Time-domain simulations use the operating profile in Fig.~\ref{fg:case2}(c), including smooth changes in PV power targets and step changes in CPL demand targets. All PI and \revB{OFP-AD} gains remain fixed. The node voltages settle after these changes (Fig.~\ref{fg:case2}(d)), illustrating small-signal stability along the prescribed operating set.

\begin{figure}[t]
    \centering
\includegraphics[width=0.8\linewidth]{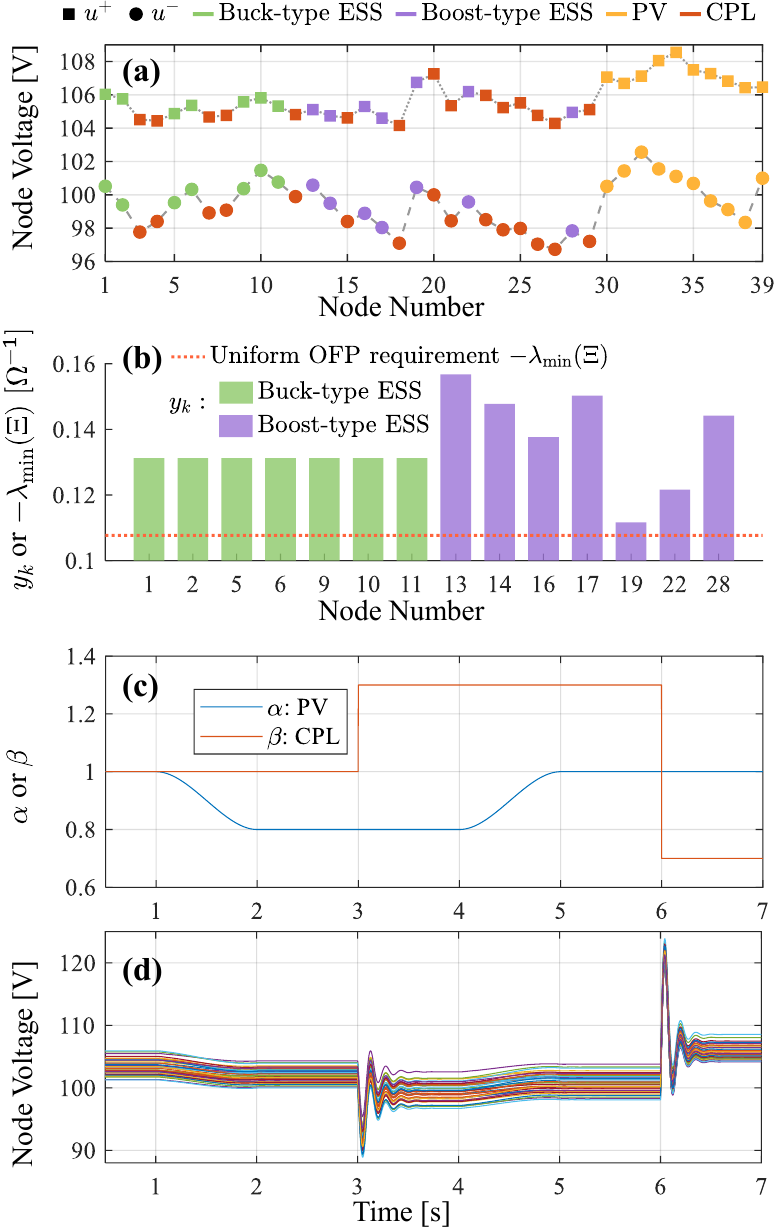}
\vspace{-0.5\baselineskip}
    \caption{Case 2: (a) equilibrium-voltage bounds; (b) minimum ESS OFP indices over sampled voltages and the uniform OFP requirement; (c) PV and CPL power-scaling profiles; (d) node-voltage responses.}
    \label{fg:case2}
    \vspace{-1\baselineskip}
\end{figure}

\section{Conclusion}
The proposed passivity-based decentralized stability criterion
\revB{accommodates coexisting fixed- and tunable-controller devices and}
provides a sufficient certificate of DCMG
small-signal stability. 
The full-order analytical bounds show that \revB{finite-bandwidth current loops limit the OFP indices of droop-controlled converters. Consequently,} positive
OFP indices predicted by reduced-order models
cannot always be obtained through PI gain tuning.
The proposed \revB{OFP-AD can improve} OFP indices while retaining cascaded PI control and \revB{the original droop equilibrium}. 
\revA{The criterion specifies quantitative OFP requirements that each device can verify locally without disclosing its model.} \revD{The results can thus support the formulation of stability provisions in DCMG grid codes, as well as device-level active-damping design and parameter tuning. Future work will address coordinated allocation of heterogeneous OFP requirements and the effects of digital control delays on device OFP indices.}

\appendices
\refstepcounter{section}\label{ap:Equilibrium-Voltage Bounds}
\section*{Appendix~\thesection: Equilibrium-Voltage Bounds}
At steady state, each droop-controlled ESS acts as a voltage source $U_k^{\mathrm r}$ behind a resistance $R_k^{\mathrm d}$, so the results of \cite{jeeninga2022dcPartI,jeeninga2022dcPartII} apply to the equilibrium equations $Y_\mrm{net}\bm u=\bm i(\bm u)$, with $\bm i(\bm u)$ the steady-state injection vector. 
An equilibrium is long-term voltage-stable, a static property of the power flow \cite[Def.~2.8]{jeeninga2022dcPartI}, if and only if $J=Y_\mrm{net}-\partial\bm i/\partial\bm u \succ0$ \cite[Prop.~3.3]{jeeninga2022dcPartI}. 
Since $\lambda_{\min}(J)=0.5298>0$ at the max-demand equilibrium $\bm u^-$, this equilibrium is long-term voltage-stable and its operating condition lies strictly within the loadability limit \cite[Cor.~3.21]{jeeninga2022dcPartI}.
So does every other operating condition in the prescribed range, since its net power demand is componentwise no greater \cite[Lem.~7.1]{jeeninga2022dcPartII}. Each thus admits a unique long-term voltage-stable solution \cite[Cor.~3.21]{jeeninga2022dcPartI}, namely the high-voltage solution \cite[Th.~8.1]{jeeninga2022dcPartII}. 
Along this solution branch, the voltages are nondecreasing in PV generation and nonincreasing in load demand \cite[Lem.~3.16 and Prop.~3.3]{jeeninga2022dcPartI}. Thus, $\bm u^-$ and $\bm u^+$ bound the high-voltage solutions of all intermediate operating conditions as in \eqref{eq:case2_voltage_bounds}.

\ifCLASSOPTIONcaptionsoff
  \newpage
\fi





\bibliographystyle{IEEEtran}
\bibliography{IEEEabrv,Bibliography}
%





\vfill


\end{document}